\documentclass[submission,Phys]{SciPost}

\usepackage{amsmath}
\usepackage{graphicx}
\usepackage{amsfonts}
\usepackage{amssymb}

\usepackage{color}
\usepackage{dsfont}
\usepackage{amsmath} 
\usepackage{hyperref}
\usepackage{physics}
\usepackage{orcidlink}
\hypersetup{colorlinks = true, urlcolor = blue, linkcolor = blue, citecolor = blue}

\newcounter{CommentNumber}
\newcommand{\comment}[1]{\stepcounter{CommentNumber}\belowpdfbookmark{#1}{\arabic{CommentNumber}}}

\begin{document}

\begin{center}{\Large \textbf{
 Pareto-optimality of Majoranas in hybrid platforms
}}\end{center}

\begin{center}
Juan Daniel Torres Luna\textsuperscript{1, 2}\orcidlink{0009-0001-6542-6518},
Sebastian Miles\textsuperscript{1, 2}\orcidlink{0009-0005-6425-8072},
A. Mert Bozkurt\textsuperscript{1, 2}\orcidlink{0000-0003-0593-6062},
Chun-Xiao Liu\textsuperscript{1, 2}\orcidlink{0000-0002-4071-9058},
Antonio L. R. Manesco\textsuperscript{2}\orcidlink{0000-0001-7667-6283},
Anton R. Akhmerov\textsuperscript{2,*}\orcidlink{0000-0001-8031-1340},
Michael Wimmer\textsuperscript{1, 2, $\dagger$}\orcidlink{0000-0001-6654-2310}
\end{center}

\begin{center}
 {\bf 1} Qutech, Delft University of Technology, Delft 2600 GA, The Netherlands
    \\
 {\bf 2} Kavli Institute of Nanoscience, Delft University of Technology, Delft 2600 GA, The Netherlands
    \\
    \textsuperscript{$*$}\href{mailto:pareto\_majoranas@antonakhmerov.org}{pareto\_majoranas@antonakhmerov.org},\textsuperscript{$\dag$}\href{mailto:m.t.wimmer@tudelft.nl}{m.t.wimmer@tudelft.nl}.
\end{center}
\begin{center}
October 08, 2025
\end{center}

\begin{abstract}
To observe Majorana bound states, and especially to use them as a qubit, requires careful optimization of competing quality metrics.
We systematically compare Majorana quality in proximitized semiconductor nanowires and quantum dot chains.
Using multi-objective optimization, we analyze the fundamental trade-offs between topological gap and localization length—two key metrics that determine MBS coherence and operational fidelity.
We demonstrate that these quantities cannot be simultaneously optimized in realistic models, creating Pareto frontiers that define the achievable parameter space.
Our results show that QD chains achieve both comparable quality as nanowires and a regime with a much shorter localization length, making them particularly promising for near-term quantum computing applications where device length and disorder are limiting factors.
\end{abstract}

\section{Introduction}

\comment{Majorana qubits are interesting because their topological protection exponentially suppresses local errors, but this protection requires several conditions to be fulfilled.}
Majorana bound states (MBS) are topologically protected states that appear at the boundary of one-dimensional topological superconductors~\cite{Kitaev2001Unpaired}.
Two spatially separated MBS encode a single fermion protected against local perturbations, making them resilient against local sources of disorder~\cite{Boross2024Braiding,Sau2012,Plugge2017Majorana}.
Their non-local nature allows for topologically protected operations---braiding---within the degenerate manifold of MBS.
Consequently, experimental and theoretical researchers pursue MBS to realize Majorana qubits~\cite{Aghaee2023InAs,Aghaee2024Interferometric,Zatelli2023Robust,Bordin2023Tunable,Bordin2024Crossed,Wu2021Triple,Bordin2024Signatures,vanLoo2025}.

\comment{The quality of Majorana qubits is limited by the topological gap and the localization length.}
Creating a Majorana qubit and demonstrating MBS non-Abelian exchange statistics requires optimization of the error-rate per braiding operation~\cite{vanHeck2012,Aasen2016Milestones,Plugge2017Majorana,tsintzis_majorana_2024,MS_roadmap_2025}.
On the one hand, braiding fails if it is fast enough to excite non-equilibrium quasiparticles above the topological gap $E_\text{gap}$---the energy difference between the ground state and the first excited state in the topological phase.
On the other hand, the dephasing timescale of a Majorana qubit depends on the overlap between the two separate MBS that extend over a region of size $\xi$---the localization length.
Optimizing the topological gap and the localization length brings a Majorana qubit to the desired regime where quantum information is protected against decoherence in a timescale appropriate for performing braiding.

\comment{There are two state-of-the-art approaches to realizing Majorana bound states in hybrid semiconducting devices.}
Proximitized semiconducting nanowires with spin-orbit coupling and a magnetic field are one of the first hybrid systems predicted to host Majorana bound states~\cite{Lutchyn2010Majorana,Oreg2010Helical}.
In this work, we refer to this model as the nanowire (NW) model.
In the clean limit, the system reaches the topological phase by tuning the electrostatic potential along the wire and the magnetic field~\cite{Lutchyn2011Search,Pikulin2012,SauDasSarma2013}.
Despite efforts to mitigate strong disorder in long semiconductor nanowires~\cite{Aghaee2023InAs}, current devices do not meet the requirements for high-quality MBS~\cite{MS_XZ_2025}.
Quantum dot (QD) chains provide a bottom-up approach where alternating normal and proximitized dots~\cite{Sau2012,Leijnse2012Parity,Liu2022Tunable,Tsintzis2022Creating} create an effective Kitaev chain~\cite{Kitaev2001Unpaired} with tunable parameters.
Although the QD chain approach enables tuning away disorder, optimally tuning and controlling a long QD chain remains a challenge.
The question about which approach will succeed in realizing scalable, high-quality Majorana qubits given the challenges of disorder and tuning remains open.

\comment{We compare the quality of Majorana bound states in two state-of-the-art platforms: the Lutchyn-Oreg model and the quantum dot chain.}
In this work we introduce a systematic approach to characterise and compare the performance of different platforms for Majorana qubits.
To this end, we reformulate the problem of characterizing a Majorana platform as a multi-objective optimization (MOO) problem where the Pareto front measures the best possible quality achievable by a given model.
Our work demonstrates that both the NW and QD models achieve similar quality of MBS in the clean limit, but the QD chain reaches a regime with very short localization length that is inaccessible to the NW model.
The content of the manuscript is organized as follows:
In Sec.~\ref{sec:quality} we discuss the requirements for obtaining high-quality MBS.
In Sec.~\ref{sec:NW_model} we study the quality of MBS in the NW model.
In Sec.~\ref{sec:qd_chain} we use perturbation theory to understand the weak-coupling limit of the QD chain.
We also discuss a protocol for optimally tuning QD chains.
In Sec.~\ref{sec:comparison} we perform multi-objective optimization to compare both models in the clean limit.
We conclude in Sec.~\ref{sec:conclusions} by highlighting the experimental implications of our work for near-term experiments.

\section{Quality metrics for Majorana qubits}\label{sec:quality}

\comment{Topological qubits are protected against local errors, but this protection requires several conditions to be fulfilled.}
Majorana-based quantum computing relies on the non-Abelian exchange statistics of MBS to perform protected quantum operations---braiding.
To successfully braid MBS in a real device, we must consider the following requirements~\cite{vanHeck2012, Plugge2017Majorana,tsintzis_majorana_2024,MS_roadmap_2025}:
\begin{itemize}
\item MBS must be sufficiently far apart so that they are effectively decoupled,
\item the ground state parity must not change due to thermal excitations, and
\item manipulations must be adiabatic to avoid Landau-Zener excitations.
\end{itemize}
Each requirement corresponds to a different physical mechanism and introduces a characteristic time scale that must be considered when designing a braiding protocol.
First, in a finite system, the overlap of MBS results in a finite coupling strength, which defines a time scale $t_M \sim \exp(-L/\xi)$, where $L$ is the system size and $\xi$ is the localization length of the MBS.
Second, the ground state parity flips due to quasiparticle poisoning events from thermally excited states in the superconductor.
These events occur in a time scale $t_\text{qp} \sim \exp(E_\text{gap}/k_B T)$, where $E_\text{gap}$ is the topological gap, $k_B$ is the Boltzmann constant, and $T$ is the temperature.
Finally, to avoid non-adiabatic transitions, the braiding operation must be much slower than the Landau-Zener time scale $t_\text{LZ} \sim \hbar/E_\text{gap}$.
Satisfying all conditions requires the following inequality to hold:
\begin{equation}\label{eq:inequality}
t_\text{LZ} \ll t_\text{braid} \ll t_M, t_\text{qp}.
\end{equation}
Therefore, the optimal regime for operating a Majorana qubit involves tuning the microscopic parameters $\mathbf{x}$ so that topological gap $E_\text{gap}(\mathbf{x})$ is maximal and the localization length $\xi(\mathbf{x})$ is minimal.
These conditions together define a multi-objective optimization problem where two competing objectives must be optimized simultaneously.

\comment{The error rate per braiding operation depends on intrinsic model parameters and extrinsic device parameters.}
Successfully braiding MBS requires to optimise the error-rate per braiding operation, $\Gamma(\xi, E_\text{gap}, L, T)$, for a particular realization of a device.
This error rate depends both on intrinsic properties of the MBS, namely the topological gap $E_\text{gap}$ and the localization length $\xi$, and extrinsic device parameters such as the temperature $T$ and the system size $L$.
By tuning both microscopic parameters and device design parameters, we aim to minimize $\Gamma$ and to find the optimal error rate per braiding operation
\begin{equation}\label{eq:optimal_error}
\Gamma_\text{opt}(L, T) = \min_{\mathbf{x}} \Gamma(\xi(\mathbf{x}), E_\text{gap}(\mathbf{x}), L, T).
\end{equation}

\begin{figure}[h!]
  \centering
  \includegraphics[width=0.8\linewidth]{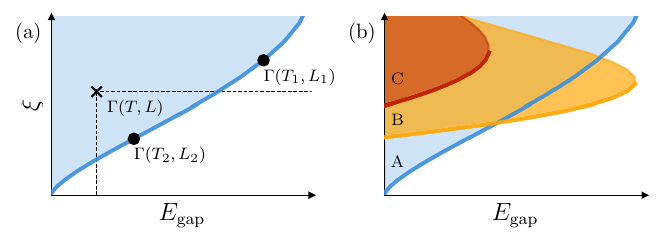}
  \caption{
    Illustration of the MOO problem and the Pareto front.
    The filled regions represent all the possible combinations of $E_\text{gap}$ and $\xi$ for a given platform, and the solid lines represent the Pareto front.
    (a) A single platform's Pareto front with two solutions with an optimal error rate for two choices of $(T_1, L_1)$ and $(T_2, L_2)$.
    The dashed lines indicate the quadrant where all solutions are strictly better for any choice of $L$ and $T$.
    (b) The Pareto fronts of three different platforms.
    }\label{fig:illustration}
\end{figure}

\comment{We use the Pareto front to choose the extrinsic device parameters.}
In this work we formulate Eq.~\eqref{eq:optimal_error} as a multi-objective optimization problem and use the Pareto front as a tool to identify the optimal operating points of a given platform.
We illustrate this idea in Fig.~\ref{fig:illustration} (a).
By changing the microscopic parameters $\mathbf{x}$, we find all possible combinations of $E_\text{gap}$ and $\xi$ that a given model can achieve.
The boundary with a positive slope (solid red line) represents the set of optimal trade-offs, known as the Pareto front, where improving one quantity necessarily worsens the other.
The point $\Gamma(T, L)$ is a suboptimal solution because it has a strictly worse error rate than all solutions in the quadrant defined by the dashed lines.
On the other hand, the two points $\Gamma(T_1, L_1)$ and $\Gamma(T_2, L_2)$ are optimal error rates for different choices of extrinsic parameters $T$ and $L$.

\comment{We use the Pareto front to compare different platforms for realizing Majorana qubits.}
Besides finding the optimal point of a given platform, we develop a systematic way to compare different platforms for realizing Majorana qubits using the Pareto front.
In Fig.~\ref{fig:illustration} (b), we illustrate the Pareto front of three different platforms.
We observe that platform C is strictly worse than both platforms A and B since it always has a higher error rate $\Gamma_\text{braid}$ than the other two platforms.
On the other hand, platforms A and B reach different regions of the space $(\xi, E_\text{gap})$ and are therefore optimal for different choices of extrinsic parameters $T$ and $L$.
For example, platform A is strictly better than platform B in the regime with very small $\xi$ and small $E_\text{gap}$, which is ideal for short devices at very low temperatures.
On the other hand, platform B is strictly better than platform A in the regime of large $E_\text{gap}$ and large $\xi$, which is ideal for long devices at higher temperatures.
In practice, we use the package \texttt{pymoo}~\cite{pymoo} to calculate the Pareto front of a given model.


\comment{In the Kitaev chain, the Pareto front is a single point where the topological gap is maximized and the localization length is zero.}
We first analyze the simplest case of a Majorana system: the Kitaev chain, where both the gap and the localization length can be optimized simultaneously.
The Kitaev chain~\cite{Kitaev2001Unpaired} is a minimal model for MBS that captures the essential physics of topological superconductivity.
The model corresponds to the Hamiltonian
\begin{equation}\label{eq:Hkitaev}
  H_\text{Kitaev} = -\sum_{j=1}^{L-1} \left( t c_j^\dagger c_{j+1} + \Delta_p c_j^\dagger c_{j+1}^\dagger + \text{h.c.} \right) - \mu \sum_{j=1}^L c_j^\dagger c_j,
\end{equation}
where $c_j$ is the annihilation operator for a fermion at site $j$, $t$ is the hopping amplitude, $\Delta_p$ is the $p$-wave pairing amplitude, and $\mu$ is the chemical potential.
MBS exist throughout the topological phase, but at the so-called sweet spot, $t=\Delta_p$ and $\mu=0$, the topological gap is maximized and the localization length is zero.
In this case, the Pareto front is a single point where the topological gap is maximized and the localization length is zero.
Realistic platforms do not exhibit this ideal situation, and the topological gap and the localization length inter-depend on the microscopic details of the system~\cite{Zatelli2023Robust,Aghaee2023InAs}.

\section{Majorana quality in the nanowire model} \label{sec:NW_model}

\begin{figure}[h!]
  \centering
  \includegraphics[width=\linewidth]{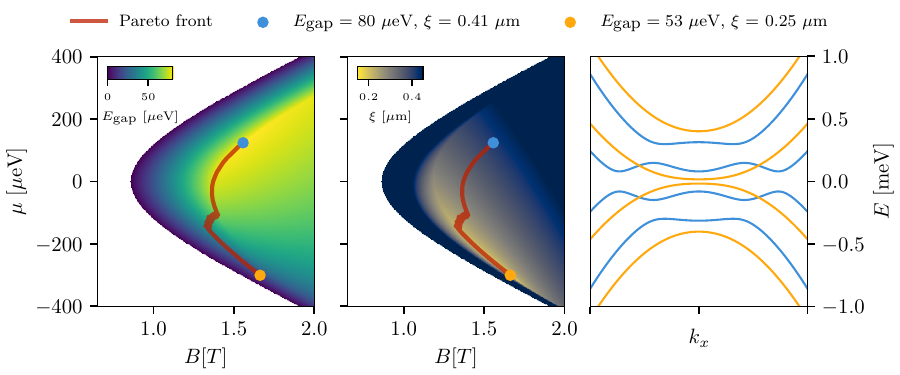}
  \caption{
    Majorana quality in the topological phase of the NW model given by Eq.~\eqref{eq:HNW}.
    (a) Topological gap $E_\text{gap}$ as a function of the chemical potential $\mu$ and the magnetic field $B$.
    (b) Localization length $\xi$ as a function of the chemical potential $\mu$ and the magnetic field $B$.
    The red line is the Pareto front, and its end points---the orange and blue dots are the parameters that optimize $\xi$ or $E_\text{gap}$, respectively.
    (c) Band structure of the two extreme cases given by the orange and blue dots in panels (a) and (b).
    We use the parameters listed in Table~\ref{tab:params} as $\text{InSbAs}$ with $m_\text{eff}/m_e=0.0162$, $g_\text{sm}=6.8$, $\alpha=20$ nm meV, and $\Delta=0.2$ meV.
    }\label{fig:MS_PD}
\end{figure}

\comment{The Lutchyn-Oreg model is the most widespread model to find Majoranas in semiconducting devices.}
We simulate a clean hybrid nanowire of length $L$~\cite{Lutchyn2010Majorana,Oreg2010Helical} described by the Hamiltonian
\begin{align}
  H_\text{NW} &= H_\text{sm} + H_\text{sc},\label{eq:HNW}\\
  H_\text{sm} &= \sum_{\sigma\sigma'} \int_0^L dx \psi_\sigma^\dagger(x) \left( -\frac{\hbar^2}{2 m_\text{eff}}\partial_x^2 + i \sigma_y \alpha \partial_x + E_Z \sigma_x \right) \psi_{\sigma'}(x),\label{eq:Hsm}\\
  H_\text{sc} &= \Delta \int_0^L dx \left( \psi_{\uparrow}(x)^\dagger\psi_{\downarrow}(x)^\dagger +\text{h.c.} \right),\label{eq:Hsc}
\end{align}
where $\psi_\sigma(x)$ is the annihilation operator for an electron with spin $\sigma$ at position $x$ in the nanowire, $m_\text{eff}$ is the effective mass, $\alpha$ is the Rashba spin-orbit coupling, $E_Z=\mu_B g_\text{sm} B /2$ is the Zeeman energy where $\mu_B$ is the Bohr magneton and $g_\text{sm}$ is the semiconducting g-factor and $B$ is the magnetic field, and $\Delta$ is the induced superconducting pairing amplitude.
The system undergoes a phase transition into the topological phase when the magnetic field satisfies $E_Z > \sqrt{\Delta^2 + \mu^2}$~\cite{Lutchyn2010Majorana,Oreg2010Helical}.
In this phase, two MBS are located at the ends of the nanowire with zero energy.

\comment{To determine the quality of the MBS, we compute the localization length and the gap.}
To find the optimal regime, we compute the topological gap $E_\text{gap}$ and the localization length $\xi$ for different values of $E_Z$ and $\mu$ in an infinite system using the momentum space Bogoliubov-de Gennes (BdG) representation of the Hamiltonian in Eq.~\eqref{eq:HNW} using Kwant package~\cite{kwant}.
To find the topological gap, we use a binary search method of the propagating modes as done in Ref.~\cite{Nijholt2016Orbital}.
The localization length is
\begin{equation}\label{eq:localization_length}
  \xi = -\frac{1}{\log(\lambda_\text{max})},
\end{equation}
where $\lambda_\text{max}$ is the eigenvalue of the transfer matrix associated with the slowest decaying mode, that is, the eigenvalue with the largest absolute value inside the unit circle.

\comment{These two quantities, however, are not independent and cannot be simultaneously optimized.}
In Fig.~\ref{fig:MS_PD} we show the topological gap (a) and the localization length (b) inside of the topological phase as a function of magnetic field $B$ and chemical potential $\mu$.
In contrast to the Kitaev chain~\cite{Kitaev2001Unpaired}, we find that in the NW model, the global maximum of the topological gap $E_\text{gap}$ differs from the global minimum of the localization length $\xi$.
The highest quality MBS appear along two distinct ridges in parameter space, where $E_\text{gap}$ and $\xi$ have global maxima and minima, respectively, as shown in the red region in Fig.~\ref{fig:MS_PD}~(a) and (b).
The ridge for positive $\mu$ maximizes $E_\text{gap}$ since it has a large Fermi wave-vector $k_F$, and thus a large spin-orbit energy $\alpha k_F$.
Therefore, it opens two topological gaps as shown by the blue line in Fig.~\ref{fig:MS_PD}~(c).
Minimizing $\xi$, on the other hand, requires a small Fermi velocity $v_F$ which occurs at the bottom of the band where $\mu$ is negative with a single gap opening around $k_x\sim0$ as shown by the orange line in Fig.~\ref{fig:MS_PD}~(c).
This creates a fundamental trade-off: achieving large gaps requires positive $\mu$ and large $k_F$, while achieving short localization lengths requires negative $\mu$ and small $v_F$.

\section{Majorana quality in the quantum dot chain}\label{sec:qd_chain}

\begin{figure}[h!]
  \centering
  \includegraphics[width=0.8\linewidth]{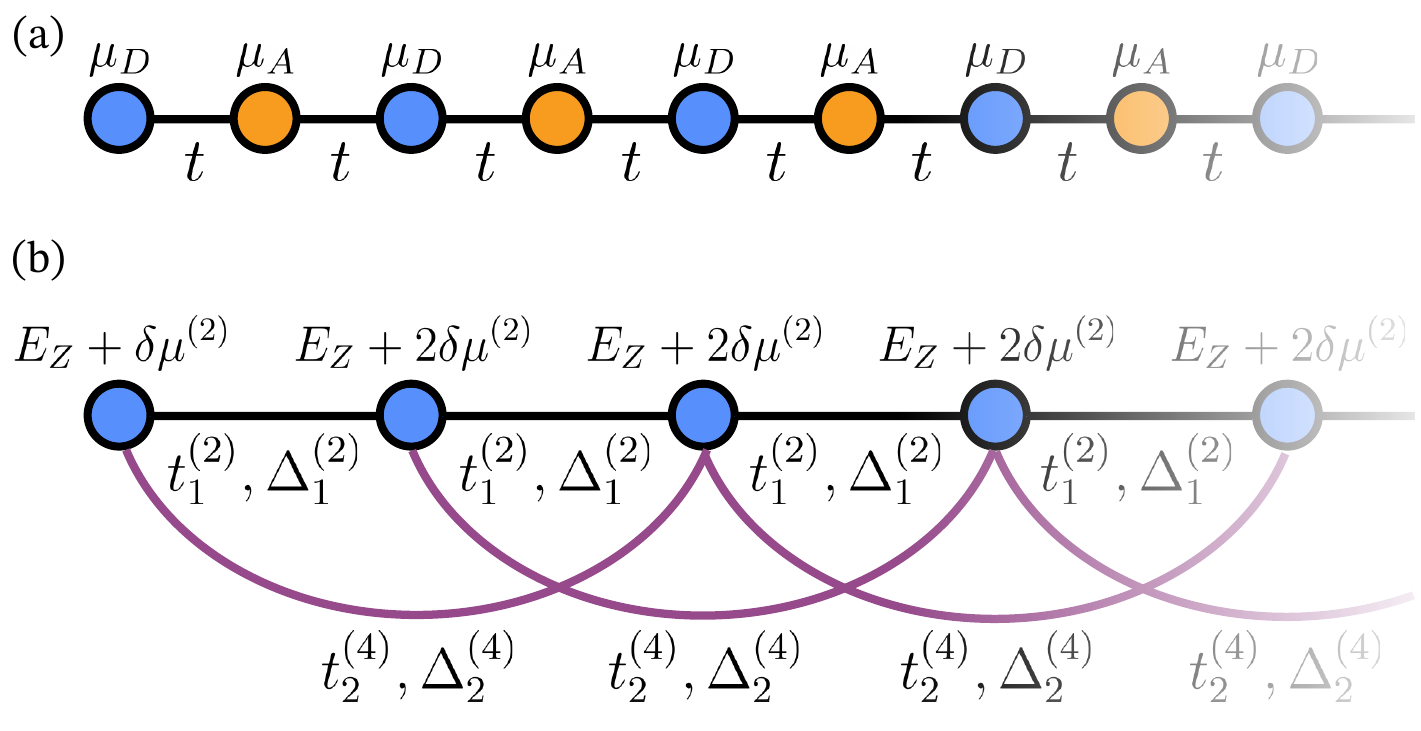}
  \caption{
    (a) Schematic of the quantum dot chain with $N$ quantum dots (blue) connected by $N-1$ ABS (orange) via hopping with amplitude $t$.
    Quantum dots have chemical potential $\mu_D$ and ABS have chemical potential $\mu_A$.
    (b) Schematic of the effective Hamiltonian for the quantum dot chain with position-dependent chemical potential, nearest-neighbor CAR and ECT, and long-range couplings (purple lines).
    }\label{fig:chain}
\end{figure}

\comment{To overcome the limitations of the nanowire model, the quantum dot chain emerges as a promising bottom-up approach to realize Majoranas.}
To overcome the disorder sensitivity of long nanowires~\cite{Aghaee2023InAs,DasSarma2016How}, building a QD chain~\cite{Sau2012,Leijnse2012Parity,Fulga2013Adaptive,Liu2022Tunable,Tsintzis2022Creating} offers control over the disorder profile along the chain, providing an alternative approach to realize MBS.
To study the quality of MBS in the QD chain, we consider the Hamiltonian of a chain of $N$ quantum dots connected by $N-1$ ABS as shown in Fig.~\ref{fig:chain}(a).
For simplicity, we assume uniform parameters across the chain.
The Hamiltonian that describes this system is
\begin{align}
  H &= H_{\text{dots}} + H_{\text{ABS}} + H_{\text{tun}} + H_{\text{tun}}^\text{so},\label{eq:h_dot_abs}\\
  H_{\text{dots}} &= \sum_{i=1}^N \left[ \left(\mu_D + E_Z \right)n_{i\uparrow} + (\mu_D - E_Z) n_{i\downarrow}\right],\\
  H_{\text{ABS}} &= \sum_{i=1}^{N-1} \left[  (\mu_A + E_Z^\text{ABS}) \tilde n_{i\uparrow} + (\mu_A - E_Z^\text{ABS})\tilde n_{i\downarrow} + \Delta (d_{i\uparrow}^\dagger d_{i\downarrow}^\dagger - d_{i\downarrow}^\dagger d_{i\uparrow}^\dagger)\right] + \text{h.c.},\label{eq:HABS}\\
  H_{\text{tun}} &=  t \cos\left(\frac{\theta}{2}\right) \sum_{i=1}^{N-1} \left[c_{i\uparrow}^\dagger d_{i\uparrow} + d_{i\uparrow}^\dagger c_{i + 1\uparrow} + c_{i\downarrow}^\dagger d_{i\downarrow} + d_{i\downarrow}^\dagger c_{i + 1\downarrow} \right] + \text{h.c.},\\
  H_{\text{tun}}^\text{so} &= t  \sin\left(\frac{\theta}{2}\right) \sum_{i=1}^{N-1} \left[c_{i\uparrow}^\dagger d_{i\downarrow} + d_{i\uparrow}^\dagger c_{i + 1\downarrow} - c_{i\downarrow}^\dagger d_{i\uparrow} -  d_{i\downarrow}^\dagger c_{i + 1\uparrow} \right] +\text{h.c.},
\end{align}
where $c_{i\sigma}$ and $d_{i\sigma}$ are the annihilation operators for the quantum dot and the Andreev bound state at site $i$, respectively.
The number operator of a quantum dot with spin $\sigma$ at site $i$ is $n_{i\sigma} = c_{i\sigma}^\dagger c_{i\sigma}$.
Similarly, the number operator for the ABS is $\tilde n_{i\sigma} = d_{i\sigma}^\dagger d_{i\sigma}$.
Here $\mu_D$ is the chemical potential for the quantum dots, $\mu_A$ is the chemical potential of the ABS, and $\Delta$ is the superconducting pairing amplitude in the ABS.
The tunneling strength between quantum dots and ABS is $t$ and $\theta$ is the spin-orbit angle.
The Zeeman energy in the normal dots is $E_Z = \mu_B g_\text{sm} B/2$ where $\mu_B$ is the Bohr magneton, $g_\text{sm}$ is the semiconducting $g$-factor and $B$ is the magnetic field.
We also include a Zeeman energy in the ABS $E_Z^\text{ABS} = \mu_B g_\text{abs} B/2$ where $g_\text{abs}$ is the $g$-factor of the ABS.

\comment{This model assumes that the spin splitting is large enough to consider a single spin species.}
Describing a QD chain using a single-particle model relies on several assumptions.
First, in the recent experiments \cite{Zatelli2023Robust,Bordin2023Tunable,Bordin2024Signatures,vanLoo2025}, local Coulomb interaction was shown to be the largest energy scale.
When the Zeeman splitting is large enough such that effectively only a single spin species contributes to the physics of the chain, we can disregard the local Coulomb interaction.
Therefore, we use the spin expectation value on the quantum dots as a proxy for when the model is a valid representation of experiments.
For a more extensive discussion, we refer the reader to Appendix~\ref{app:distribution}.
Second, we do not include the continuum of states in the superconductor, therefore we require the ABS energy to be within the induced gap.
Finally, for simplicity we start the discussion assuming that $g_\text{ABS}=0$, and later relax this condition~\cite{Vilkelis2024Fermionic}.

\comment{We obtain an effective Hamiltonian for a spinless quantum dot chain using perturbation theory.}
The Kitaev chain~\cite{Kitaev2001Unpaired} in Eq.~\eqref{eq:Hkitaev} has perfectly localized MBS at the sweet-spot.
The QD chain achieves the Kitaev regime in the weak-coupling limit up to leading-order in perturbation theory~\cite{Liu2022Tunable}.
When $t$ increases, higher-order terms contribute to the effective Hamiltonian and the MBS localization length increases.
The weak-coupling limit is
\begin{equation}
  t \ll E_\text{ABS} \ll E_Z
\end{equation}
where $E_\text{ABS} = \sqrt{\mu_A^2 + \Delta^2}$ is the ABS energy.
We perform a Schrieffer-Wolff transformation~\cite{Day2024Pymablock,Day2024Pymablock_2} of Eq.~\eqref{eq:h_dot_abs} into the regime where all QDs are spin polarized, and we illustrate the effective Hamiltonian up to fourth-order in Fig.~\ref{fig:chain}(b).
The effective Hamiltonian is
\begin{align}
  \mathcal{H} &= \sum_{i=1}^N \mu_i c_{i}^\dagger c_i  + \mathcal{H}^{(2)} + \mathcal{H}^{(4)},\label{eq:H_dot_abs}
\end{align}
where $c_i \equiv c_{i\downarrow}$ is the annihilation operator for the quantum dot at site $i$, $\mu_i=\mu_D - E_Z$ is the chemical potential tuned to the spin-up dot level, $\mathcal{H}^{(2)}$ is the second-order effective Hamiltonian, and $\mathcal{H}^{(4)}$ is the fourth-order effective Hamiltonian.

\subsection{Second-order effective Hamiltonian}\label{sec:second_order}
The second order effective Hamiltonian is
\begin{align}
  \mathcal{H}^{(2)} = t^2\left[ \sum_{i=1}^N \delta\mu_i^{(2)} c_{i}^\dagger c_i + \sum_{i=1}^{N-1} t_1^{(2)} c_i^\dagger c_{i+1} + \Delta_1^{(2)} c_i^\dagger c_{i+1}^\dagger \right] +\text{h.c.}\label{eq:H2},
\end{align}
where $\delta\mu_i^{(2)}$ is the effective chemical potential at site $i$, $t_1^{(2)}$ is the nearest-neighbor elastic co-tunneling (ECT) amplitude, and $\Delta_1^{(2)}$ is the nearest-neighbor crossed Andreev reflection (CAR) amplitude.
To clarify the role of different terms and orders, we make the dependence on $t$ explicit.
The analytical expressions for the effective parameters are given in Appendix~\ref{app:effective_model}.

\begin{figure}[h!]
  \centering
  \includegraphics[width=0.85\linewidth]{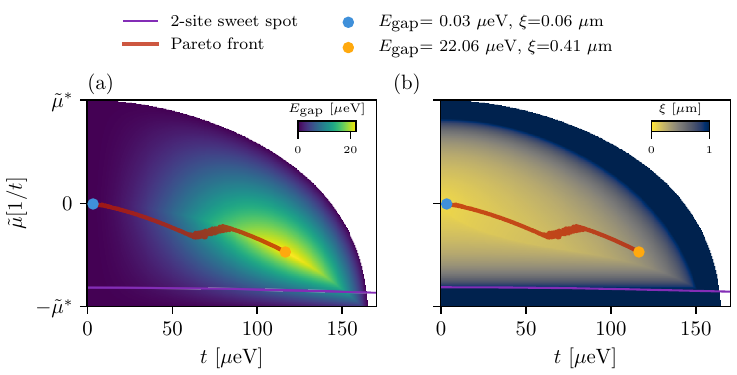}
  \caption{
    Majorana quality in the topological phase of the QD chain given by Eq.~\eqref{eq:h_dot_abs} in the weak-coupling limit.
    (a) Topological gap $E_\text{gap}$ as a function of the coupling strength $t$ and the rescaled and recentered chemical potential $\tilde{\mu}$ such that $\mu_D = (\tilde{\mu} + \delta\mu^{(2)})t^2$.
    (b) Localization length $\xi$ as a function of the coupling strength $t$ and the rescaled and recentered chemical potential $\tilde{\mu}$.
    Red region in panels (a) and (b) show the Pareto front for the QD chain~\cite{pymoo}.
    The blue and orange dots indicate the parameters that optimize $\xi$ or $E_\text{gap}$, respectively.
    The purple line indicates the two-site sweet spot condition $\mu_D = E_Z$.
    }\label{fig:QD_PD}
\end{figure}

\comment{QD chain reproduces the sweet spot of Kitaev chain to 2nd order in the interdot hopping.}
At the sweet spot $t_1^{(2)}=\Delta_1^{(2)}$, this model almost recovers the uniform Kitaev chain~\cite{Kitaev2001Unpaired}.
However, there is a crucial difference due to the presence of renormalized edge potential.
Because sites at the edges connect to only one ABS, whereas sites in the bulk connect to two ABS, the bulk sites have twice the renormalization of the edge sites.
Consequently, the effective chemical potential is
\begin{equation}\label{eq:delta_mu}
  t^2 \delta\mu_i^{(2)} = \frac{t^2 \mu_{A}}{\Delta^{2} + \mu_{A}^{2}}. \left( \delta_{i1} + \delta_{iN} + 2 \sum_{j=2}^{N-1} \delta_{ij} \right)
\end{equation}
where $\delta_{ij}$ is the Kronecker delta.
A uniform Kitaev chain enters the topological regime when $|\mu_i| < 2 |t_1^{(2)}| = 2|\Delta_1^{(2)}|$~\cite{Kitaev2001Unpaired}.
Inside this phase, the optimal gap and localization occur at the two-site sweet spot where $\mu_i=\mu = 0$ and $t_1^{(2)}=\Delta_1^{(2)}$.
For a two-site chain, the two QDs are connected to a single ABS and therefore they acquire the same renormalization.
When extending the two-site optimal parameters to a long QD chain as in Ref.~\cite{Luethi2024Fate,Luethi2025Prevalence}, the system remains topological only when the renormalization of the edges is smaller than the gap, that is, $|\delta\mu^{(2)}| < 2 |t_1^{(2)}|$.
We find that this condition constraints the spin-orbit angle to be $|\theta| < \pi/3$.
If the $\theta$ does not satisfy this condition, the edge chemical potentials must be adjusted for the system to remain topological.
Otherwise the two-site sweet spot corresponds to the trivial phase (a regime dubbed "false PMM" in Ref.~\cite{Luethi2024Fate} and Ref.~\cite{Luethi2025Prevalence}).
In Fig.~\ref{fig:QD_PD} we illustrate how the two-site sweet spot (purple line) is not optimal and does not always stays inside the topological phase.

\subsection{Fourth-order effective Hamiltonian}
\comment{The fourth-order effective Hamiltonian describes the long-range couplings.}
While in second order we have recovered the Kitaev chain with a renormalized potential at the sweet spot $t_1^{(2)}=\Delta_1^{(2)}$, the topological gap stay limited.
Therefore, we investigate the case of stronger coupling that increases the gap, but also inevitably leads to longer-range interactions due to higher-order processes.
The fourth-order effective Hamiltonian is
\begin{align}
  \mathcal{H}^{(4)} &= t^4\left[ \sum_{i=1}^N \delta\mu_i^{(4)} c_{i}^\dagger c_i + \sum_{i=1}^{N-2} t_2^{(4)} c_i^\dagger c_{i+2}+ \Delta_2^{(4)} c_i^\dagger c_{i+2}^\dagger\right]+\text{h.c.}\label{eq:H4},
\end{align}
where $\delta\mu_i^{(4)}$ is the effective chemical potential at site $i$, $t_2^{(4)}$ is the long-range hopping amplitude, and $\Delta_2^{(4)}$ is the long-range ECT amplitude.
We find that the structure of the fourth-order chemical potential is similar to the second-order chemical potential, but with a different prefactor.
The analytical expressions for the effective parameters are given in Appendix~\ref{app:effective_model}.

\comment{The long-range couplings increase the localization length.}
When the coupling $t$ becomes larger, long-range couplings appear, and they cause the localization length of the MBS to increase~\cite{Wade2013Majorana}.
To understand how the localization of the MBS $\xi$ changes due to a finite $t$, we compute $\xi$ using the eigenvalue decomposition of the transfer matrix at zero energy for an infinite QD chain, see Appendix~\ref{app:localization_length} for details.
We find that the lowest order in $t$ contribution to the localization length is
\begin{equation}\label{eq:analytical_xi}
  \xi = -\log(\lambda_\text{max})^{-1} = -\log\left[ t^{2/3} \left( \frac{\Delta_2^{(4)} - t_2^{(4)}}{\Delta_1^{(2)}} \right)^{1/3}\right]^{-1}.
\end{equation}
This equation shows that the localization length becomes finite when either the long-range couplings are different, $t_2^{(4)} \neq \Delta_2^{(4)}$, or when the coupling $t$ is finite.
Since the gap increases with $t$, that is, $E_\text{gap} \sim t^2$, optimizing Majorana quality in a QD chain defines a non-trivial MOO problem.

\comment{The optimal tuning deviates from the simple two-site sweet spot due to long-range couplings and edge effects.}
In Fig.~\ref{fig:QD_PD} we show the topological phase of the infinite QD chain in the weak-coupling limit.
We replace the chemical potential $\mu_D$ by the rescaled and recentered chemical potential $\tilde{\mu}= (\mu_D - \delta\mu^{(2)})/t^2$.
The second order optimal configuration corresponds to the line at $\tilde{\mu}=0$ in Fig.~\ref{fig:QD_PD}(a) and (b).
The optimal configuration, however, follows the Pareto front (red line) in Fig.~\ref{fig:QD_PD}(a) and (b).
For small $t$, the Pareto front has highly localized MBS with small gaps, but as $t$ increases, it deviates from the second order optimal condition and the localization length increases due to the emergence of long-range coupling (see Eq.~\eqref{eq:analytical_xi}).
The optimal operation regime follows this line and spreads over the two ridges that maximize $E_\text{gap}$ and minimize $\xi$ as shown in the red region in Fig.~\ref{fig:QD_PD}(a) and (b).
Figure~\ref{fig:QD_PD} demonstrates that using the two-site sweet spot condition~\cite{Luethi2024Fate} (see purple line) does not guarantee optimal performance and sometimes falls outside the topological phase.

\section{Comparison between the QD chain and the nanowire}\label{sec:comparison}
\begin{figure}
  \centering
  \includegraphics[width=0.75\linewidth]{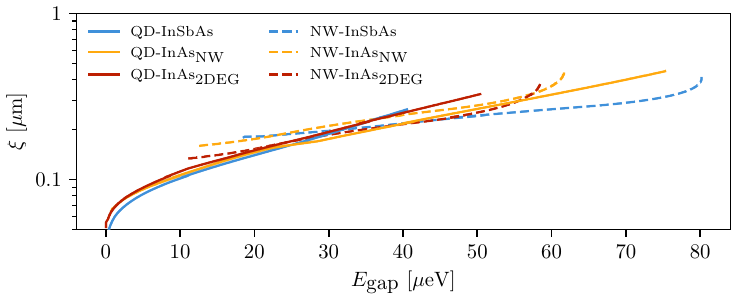}
  \caption{
    Comparison of the Pareto fronts between the QD chain (solid lines) and the NW model (dashed lines) for three different material parameter configurations listed in Table~\ref{tab:params}.
    Both models include renormalization effects as described in Appendix~\ref{app:renormalization}.
    The distribution parameters of the Pareto front and their cut-offs are discussed in detail in Appendix~\ref{app:distribution}.
    }\label{fig:pareto}
\end{figure}

\comment{In order to identify which platform offers the best trade-offs between topological gap and localization length and guide experimental efforts, we perform a comparison between the QD chain and the nanowires.}
Given recent experimental efforts to find MBS in nanowires~\cite{Aghaee2023InAs,Aghaee2024Interferometric,MS_XZ_2025} and QD chains\cite{Bordin2024Crossed,vanLoo2025,Zatelli2023Robust}, comparing the quality of MBS in these platforms is crucial for understanding how to how to perform quantum operations on Majorana bound states and eventually build a topological quantum computer.
In particular, the length of the nanowire and the number of dots in the QD chain must be large enough to ensure that the MBS are well separated.
At the same time, increasing the length of the nanowire or the number of dots in the chain is an experimental challenge due to disorder which limits the topological gap.
Therefore, we investigate how these two platforms compare in terms of the topological gap and the localization length.
In particular, we perform MOO to find the Pareto front for the models defined in Eq.~\eqref{eq:H_dot_abs} and Eq.~\eqref{eq:HNW}.
In the nanowire model there are only two optimization variables, $\mathbf{x}_\text{NW} = (\mu,B)$, and in the QD there are five optimization variables: $\mathbf{x}_\text{QD} = (\mu_D,\mu_A,t,B,\theta)$.
In App.~\ref{app:distribution} we choose the parameter ranges such that we remain in the regime of validity for both models.
Finally, we restrict ourselves to the clean limit and perfectly tuned systems in the topological phase, therefore we do not include disorder or long-range couplings in this comparison.

\comment{Since we demonstrated that the MOO problem is not trivial, we increase the complexity by including renormalization effects due to the presence of a finite magnetic field in the superconductor that proximized the ABS.}
In Sec.~\ref{sec:NW_model} and Sec.~\ref{sec:qd_chain} we found that optimizing both the topological gap and the localization length is a non-trivial problem.
In order to compare both models in a more realistic scenario, we include renormalization effects due to the presence of a finite magnetic field in the superconductor~\cite{Aghaee2023InAs} as described in Appendix~\ref{app:renormalization}.
We focus on clean systems tuned perfectly into the topological phase, therefore we disregard other factors that affect the quality of MBS, such as disorder and long-range couplings.
Furthermore, we consider a finite $g$-factor for the ABS in the QD chain, which is a realistic scenario for current experiments~\cite{Moehle2021,Bordin2023Tunable}.

\comment{We compare the results of the QD chain with the nanowire model for different material parameters.}
We present the results of the Pareto optimization in Fig.~\ref{fig:pareto} for both models and for three different material parameters listed in Table~\ref{tab:params}.
We observe that only the QD chain reaches the regime of short localization lengths, below $100$ nm, while maintaining a finite topological gap.
In this regime, the QD chain implements a true Kitaev chain, but the topological gap is small since it is proportional to the coupling between the dots and the ABS.
On the other hand, we observe that both nanowire and QD chain models overlap in the intermediate region of the Pareto front.
Both models extend over a range of $10-70$ $\mu$eV in the topological gap, while maintaining a localization length of a few hundred nanometers.

\comment{Optimizing MBS in short chains minimizes decoherence and experimental limitations.}
Device length is a crucial limiting factor in Majorana devices.
Fig.~\ref{fig:pareto} demonstrates that QD chains allow for strongly localized MBS, which in turn allows for shorter devices at the cost of a smaller topological gap.
Using a minimal distance between quantum dots of $100~$nm, we find that minimal chains of two or three sites support MBS localized in a single site.
On the other hand, nanowires have a larger localization length, which requires longer devices to ensure that the MBS are well separated.
Current nanowire devices~\cite{Aghaee2024Interferometric,MS_XZ_2025} have disorder effects that are not negligible, and therefore short QD chains may be more resilient to these effects while having strongly localized MBS.

\section{Conclusions}\label{sec:conclusions}

\comment{Our work demonstrates that the regimes where the topological gap and the localization length are optimal happen in opposite regions of parameter space.}
The topological gap and the localization length determine the quality of Majorana bound states.
We found that these two quantities cannot be simultaneously optimized in realistic models beyond the simple Kitaev model.
In the Lutchyn-Oreg model, we found that the band structure of the normal state properties defines the topological gap and the localization length.
We found that the QD chain corresponds to a Kitaev chain with long-range couplings.
The underlying structure of these models leads to a trade-off between the topological gap and the localization length.

\comment{QD chains offer distinct advantages over nanowires, particularly in achieving shorter localization lengths and disorder resilience.}
We compared the quality of Majorana bound states in two state-of-the-art platforms in the clean limit: the nanowire model and the quantum dot chain.
We found that the QD chain reaches the Kitaev limit where MBS localize in a single quantum dot.
This occurs because the QD chain can be tuned to the sweet spot of the Kitaev model, which has zero localization length and a small but finite topological gap.
Beyond this limit, both models reach a similar quality while with variations that depend on the material parameters.
We remark that we restricted ourselves to the regime where the single-particle approximation of the QD chain is valid (see Fig.~\ref{fig:spin_criteria} in Appendix~\ref{app:renormalization}).
The quality of the QD chain beyond this limit remains as an open question.
Furthermore, we expect the QD chain results to remain resilient against disorder since experimenters can tune the chain dot by dot.
The results for the NW model will be significantly affected by disorder since it cannot be tuned away as in the QD chain.
Therefore, the Pareto front for the NW model represents the best-case scenario.

\comment{The optimal tuning point of the QD chain shifts away from the two-site sweet spot condition due to the presence of long-range couplings.}
We demonstrate that the optimal tuning point of the QD chain shifts away from the two-site sweet spot condition due to the presence of long-range couplings coming from sites that are further away.
Therefore, to optimally tune a QD chain, we require an iterative procedure that starts from the two-site sweet spot condition and then re-optimizes the parameters as the length of the chain increases.
Current experiments on three-site chains have already demonstrated that the two-site sweet spot condition is not optimal since they require iterative tuning~\cite{Bordin2024Signatures,tenHaaf2025}.
Longer chains, however, will require an increasingly complex tuning procedure to reach the optimal point unless restricted to small coupling and hence small topological gaps.

\section*{Acknowledgements}
This project is supported supported by funding from Microsoft Research.
We acknowledge financial support from the Horizon Europe Framework Program of
the European Commission through the European Innovation Council Pathfinder grant No. 101115315 (QuKiT) and by the Dutch Research Council (NWO) grant OCENW.GROOT.2019.004.

\section*{Author contribution statement}
M.W. initiated the project idea on second-order perturbation theory in QD chains.
This initial project idea focused exclusively on the content described in Sec. \ref{sec:second_order}.
M.W., M.B., C.X.L. and S.M. participated in this initial project idea.
The project evolved into the results presented in the current manuscript after A.A. proposed the use of the Pareto front for comparing Majorana platforms.
J.D.T.L. was involved in all numerical and analytical calculations, result analysis, and wrote the manuscript.
M.B. contributed to the numerical and analytical calculations.
M.W. analytically calculated the Pfaffian of the QD chain.
M.W., A.A. and A.L.R.M., supervised the project.
All authors provided feedback on the manuscript and participated in analysing and discussing the results.
\section*{Data and code availability}
The code and data used to generate the results in this paper are available at Ref.~\cite{code_pmms_fate}.

\bibliography{../bibliography.bib}

\newpage
\appendix
\section{Simulation parameters}

\comment{We use experimentally relevant material parameters for three different semiconductor platforms to ensure realistic comparisons.}
In this work, we use the parameters of three different semiconductor platforms to ensure that our results are relevant for current experiments.
\begin{table}[h!]
  \centering
  \begin{tabular}{lccccccc}
    \hline
    Material & $m_{\textrm{eff}}/m_e$ & $g_{\textrm{abs}}$ & $g_{\textrm{sm}}$ & $\Delta$ [$\mu$eV] & $\Delta_\text{Al}$ [$\mu$eV] & $\alpha$ [nm$\cdot$meV] \\
    \hline
    $\text{InSbAs}$~\cite{Moehle2021} & 0.0162 & 5.5 & 6.8 & 200 & 300 & 20 \\
    $\text{InAs}_\text{NW}$~\cite{Bordin2023Tunable} & 0.023 & 21 & 40 & 260 & 300 & 11 \\
    $\text{InAs}_\text{2DEG}$~\cite{Aghaee2024Interferometric} & 0.032 & 11.4 & 11.4 & 290 & 300 & 8.3 \\
    \hline
    \end{tabular}
    \caption{
      Material parameters used in the simulations.
      The effective mass $m_{\textrm{eff}}$ is in units of the free electron mass, the $g$-factors $g_{\textrm{abs}}$ and $g_{\textrm{sm}}$ are dimensionless, $\Delta$ is the induced gap, $\Delta_\text{Al}$ is the parent superconducting gap, and $\alpha$ is the spin orbit strength.
      }\label{tab:params}
  \end{table}

\section{Analytical treatment of the QD chain}\label{app:effective_model}

\subsection{Effective Hamiltonian parameters}
\comment{The effective Hamiltonian parameters derived from perturbation theory reveal how the QD chain's behavior depends on the microscopic details.}
We perform a Schrieffer-Wolff transformation using the package \texttt{Pymablock}~\cite{Day2024Pymablock, Day2024Pymablock_2} to obtain the effective Hamiltonian parameters of the QD chain up to 4-th order in the interdot hopping $t$.
In this section we provide the resulting analytical expressions for the effective parameters of the QD chain.
While the expressions for the fourth-order chemical potential renormalization follow the same spatial structure as the second-order chemical potential renormalization in Eq.~\eqref{eq:delta_mu}, we present them in the online repository in Ref.~\cite{code_pmms_fate} since they are lengthy.
The second and fourth order nearest neighbor CAR and ECT mediated by the ABS are
\begin{align}
  t^2t_{1}^{(2)} = - \frac{\mu_{A} t^{2} \cos{\left(\theta \right)}}{\Delta^{2} + \mu_{A}^{2}}, \quad &t^2\Delta_{1}^{(2)} =- \frac{\Delta t^{2} \sin{\left(\theta \right)}}{\Delta^{2} + \mu_{A}^{2}},\\
  t^4t_{1}^{(4)}=\frac{3 t^{4} \left(2 E_{Z} \mu_{A} + \Delta^{2}\right) \cos{\left(\theta \right)}}{2 E_{Z} \left(\Delta^{4} + 2 \Delta^{2} \mu_{A}^{2} + \mu_{A}^{4}\right)}, \quad &t^4\Delta_1^{(4)} = \frac{3 \Delta t^{4} \left(E_{Z} - \mu_{A}\right) \sin{\left(\theta \right)}}{2 E_{Z} \left(\Delta^{4} + 2 \Delta^{2} \mu_{A}^{2} + \mu_{A}^{4}\right)}.
\end{align}
From the second order expressions we derive the two-site sweet spot condition, which requires the nearest neighbor CAR and ECT to be equal, that is, $t_1^{(2)} = \Delta_1^{(2)}$.
The condition for the sweet spot is
\begin{equation}\label{eq:sweet_spot}
  \mu_A = \Delta \tan(\theta).
\end{equation}
The lowest non-vanishing order for next-nearest neighbor CAR and ECT is fourth order in $t$, and they are
\begin{align}
  t^4t_2^{(4)} &= \frac{t^{4} \left[ \left(2 E_{Z} \mu_{A} + \Delta^2 \right) \cos^{2}{\left(\theta \right)}  + \mu_A^2 \sin^2\left(\theta\right)\right]}{2 E_{Z} \left(\Delta^{2} + \mu_{A}^{2}\right)^{2}},\\
  t^4\Delta_2^{(4)} &= \frac{\Delta t^{4} \left(E_{Z} - \mu_{A}\right) \sin{\left(2 \theta \right)}}{2 E_{Z} \left(\Delta^{2} + \mu_{A}^{2}\right)^2}.
\end{align}
These interactions are in general different from each other, which leads to a finite localization length as shown in Eq.~\eqref{eq:analytical_xi}.
However, in the limit $E_Z\rightarrow\infty$ and when the system is at the two-site sweet spot, they become equal, that is,
\begin{equation}
  t^4t_2^{(4)} = t^4\Delta_2^{(4)} = \frac{t^4}{\Delta^3} \sin(\theta) \cos^5(\theta).
\end{equation}
Therefore, in this limit the QD chain has perfectly localized MBS up to 4th-order.

\subsection{BdG representation of the QD chain and topological phase diagram}\label{app:topo_phase}

\comment{We write down the Hamiltonian of the QD chain the BdG representation.}
We rewrite the QD chain from Eq.~\eqref{eq:h_dot_abs} in a tight-binding representation~\cite{Luethi2024Fate},
\begin{align}\label{eq:h_dot_abs_BDG}
  H_\text{QD} &= \sum_{i=1}^{N-1} \psi_i^\dagger\left[ s_+ (\mu_D \tau_z \sigma_0 + E_Z  \tau_0 \sigma_z)  + s_-(\mu_{A} \tau_z \sigma_0 + E_Z^\text{ABS} \tau_0 \sigma_z + \Delta \tau_x \sigma_0)  \right] \psi_i \notag \\
    &+ \sum_{i=1}^{N-1} t \psi_i^\dagger s_x \tau_z (\cos (\theta/2)  \sigma_x + i \sin(\theta/2) \sigma_y ) (\psi_{i}+\psi_{i+1})+ \text{h.c.}\notag\\
    &+ \psi_N^\dagger  s_+ (\mu_D \tau_z \sigma_0 + E_Z  \tau_0 \sigma_z) \psi_N,
\end{align}
where the Nambu spinor is
\begin{equation}
    \psi_i = \left( c_{i,\uparrow}, c_{i,\downarrow}, c_{i,\downarrow}^\dagger, -c_{i,\uparrow}^\dagger, d_{i,\uparrow}, d_{i,\downarrow}, d_{i,\downarrow}^\dagger, -d_{i,\uparrow}^\dagger \right)^T,
\end{equation}
and the operators are described below Eq.~\eqref{eq:h_dot_abs}.
Here $s_i$ are the Pauli matrices for the sublattice structure, $\tau_i$ are the Pauli matrices for particle-hole space, and $\sigma_i$ are the Pauli matrices for spin.
Here $s_\pm = (s_0 \pm s_z)/2$ are the projectors to the quantum dots or the ABS subspaces.
We write this Hamiltonian in the BdG representation, which is useful to compute the topological invariant, as
\begin{equation}\label{eq:H_BdG}
    H_\text{BdG}(k) = H_\text{QD}(k) \tau_z + \Delta s_- \tau_x \sigma_0,
\end{equation}
where $H_\text{QD}(k)$ is the Fourier transform of Hamiltonian in Eq.~\eqref{eq:h_dot_abs_BDG} for $\Delta=0$, and $k$ is the one-dimensional momentum.
The spin operators are implicitly included in the Hamiltonian $H_\text{BdG}(k)$.
To compute the topological invariant, we transform Eq.~\eqref{eq:H_BdG} into an anti-symmetric form using a transformation $W$ such that $H_\mathrm{BdG}'(k) = W^\dagger H_\text{BdG}(k) W$.
The topological invariant is then
\begin{equation}
  \mathcal{Q} = \mathrm{sgn}[ \mathrm{Pf}(H_\mathrm{BdG}'(k=0)) \cdot \mathrm{Pf}(H_\mathrm{BdG}'(k=\pi))].
\end{equation}
Here, $\mathrm{Pf}$ denotes the Pfaffian.
Setting $g_\text{ABS}=0$, we find that
\begin{align}
  \mathrm{Pf}(H_\mathrm{BdG}'(k=\pi)) &= \Delta^{2} \mu_{D}^{2} + E_{Z}^{2} \left(- \Delta^{2} - \mu_{A}^{2}\right) + \left(\mu_{A} \mu_{D} - 4 t^2 \sin^{2}{\left(\frac{\theta}{2} \right)}\right)^2,\label{eq:topo_invariant_1}\\
  \mathrm{Pf}(H_\mathrm{BdG}'(k=0)) &= \Delta^{2} \mu_{D}^{2} + E_{Z}^{2} \left(- \Delta^{2} - \mu_{A}^{2}\right) + \left(\mu_{A} \mu_{D} - 4 t^2 \cos^{2}{\left(\frac{\theta}{2} \right)}\right)^2.\label{eq:topo_invariant_2}
\end{align}
Using Eq.~\eqref{eq:topo_invariant_1} and Eq.~\eqref{eq:topo_invariant_2} we find an analytical expression for the topological invariant

\comment{The topological phase diagram of the QD chain reveals multiple phases that depend on the coupling strength and chemical potential.}
\begin{figure}[h!]
  \centering
  \includegraphics[width=0.75\linewidth]{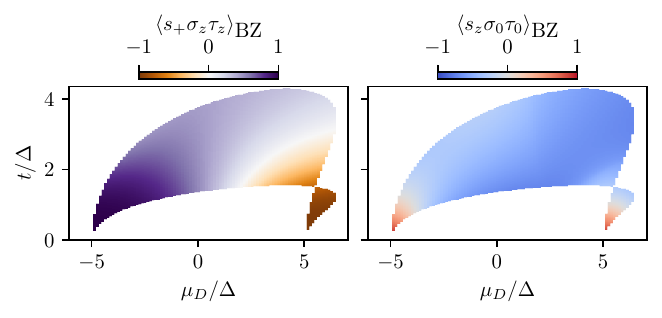}
  \caption{
    Topological phase diagram of the QD chain as a function of the coupling $t$ and the chemical potential $\mu_D$ for a fixed value of the magnetic field $E_Z/\Delta=5$, spin-orbit angle $\theta=0.21\pi$, and the ABS chemical potential $\mu_A=\Delta\tan(\theta)$ is at the sweet spot.
    We compute the average expectation value over the Brillouin zone of the renormalized spin expectation value (a) in the normal dots and the sublattice expectation value (b).
    }\label{fig:topo_phase}
\end{figure}
In Fig.~\ref{fig:topo_phase} we plot the phase diagram of the QD chain as a function of the coupling $t$ and the chemical potential $\mu_D$.
For very small $t$ we find two topological phases at $\mu_D = \pm E_Z$ with opposite spin (see Fig.~\ref{fig:topo_phase}~(a)) and separated by a trivial phase.
In the weak-coupling limit where $t \ll E_\text{ABS}$, these phases correspond to a Kitaev chain with long-range couplings.
As $t$ increases, the two topological phases merge into a single phase where the low-energy states are mostly localized in the ABS, deviating from the Kitaev chain.
The sublattice expectation value $\langle s_z \rangle$ in Fig.~\ref{fig:topo_phase}~(b) shows that the low-energy states are mostly localized in the ABS for large $t$, while they are mostly localized in the dots for small $t$.

\subsection{Calculation of the transfer matrix}\label{app:localization_length}

\comment{We compute the localization length of Majorana bound states by applying the transfer matrix methods to the effective Hamiltonian.}
To compute the localization length of the MBS in the fourth-order effective Hamiltonian shown in Eq.~\eqref{eq:H4}, we follow the method in Ref.~\cite{Antonio2017Extended}.
The BdG Hamiltonian of an infinite Kitaev chain with next-nearest-neighbor coupling is
\begin{equation}
  H_\text{BdG} = \sum_i \psi_i^\dagger \mu \tau_z \psi_i + \psi_{i+1}^\dagger \mathbf{t} \psi_i + \psi_{i+2}^\dagger \mathbf{t}' \psi_i + \text{h.c.},
\end{equation}
where $\psi_i = (c_{i},  c_{i}^\dagger)^T$ is the Nambu spinor, $\mu$ is the chemical potential, and $\mathbf{t} = t_1 \tau_z + \Delta_1 \tau_x$ and $\mathbf{t}'=t_2 \tau_z + \Delta_2 \tau_x$ are the nearest and next-nearest-neighbor hopping matrices that include CAR and ECT processes.
The Hamiltonian $H_\text{BdG}$ acting on the wavefunction at site $i$ is
\begin{equation}
  H_\text{BdG} \psi_i = \mathbf{\mu} \psi_i + \mathbf{t} \psi_{i+1} + \mathbf{t}' \psi_{i+2} + \mathbf{t} \psi_{i-1} + \mathbf{t}' \psi_{i-2}.
\end{equation}
By using particle-hole symmmetry, we can divide the transfer matrix into two blocks with symmetric solutions.
Focusing only on the zero energy solutions, we write this equation as a transfer matrix equation
\begin{equation}
  \begin{pmatrix}
    \psi_{i+2} \\
    \psi_{i+1} \\
    \psi_{i}   \\
    \psi_{i-1}
  \end{pmatrix} = T \begin{pmatrix}
    \psi_{i+1} \\
    \psi_{i} \\
    \psi_{i-1} \\
    \psi_{i-2}
  \end{pmatrix} = \begin{pmatrix}
    -\frac{\Delta_{1} + t_{1}}{\Delta_{2} + t_{2}} & - \frac{\mu}{\Delta_{2} + t_{2}} & \frac{\Delta_{1} - t_{1}}{\Delta_{2} + t_{2}} & \frac{\Delta_{2} - t_{2}}{\Delta_{2} + t_{2}}\\1 & 0 & 0 & 0\\0 & 1 & 0 & 0\\0 & 0 & 1 & 0
  \end{pmatrix} \begin{pmatrix}
    \psi_{i+1} \\
    \psi_{i} \\
    \psi_{i-1} \\
    \psi_{i-2}
  \end{pmatrix}.
\end{equation}
The eigenvalues $\lambda$ of the transfer matrix $T$ are the solution of the characteristic polynomial
\begin{equation}
  \lambda^{4} \left(\Delta_{2} + t_{2}\right) + \lambda^{3} \left(\Delta_{1} + t_{1}\right) + \lambda \left(t_{1} - \Delta_{1}\right) + (t_{2}- \Delta_{2})  = 0.
\end{equation}
We solve this polynomial and find the localization length $\xi$ of the MBS as given by Eq.~\eqref{eq:analytical_xi}.

\section{Renormalization in the microscopic model}\label{app:renormalization}

\comment{We include renormalization effects in the microscopic model due to the presence of a finite magnetic field in the superconductor that proximitsed the ABS.}
In the microscopic model, we include renormalization effects due to the presence of a finite magnetic field in the superconductor following the approach in Ref.~\cite{Aghaee2023InAs}.
That is,
\begin{equation}
  H_\text{eff} = Z^\dagger(H_\text{SM} + \Sigma_0) Z, \quad Z = (1 - \Sigma_1)^{-1/2},
\end{equation}
where $H_\text{SM}$ is the normal part of the Hamiltonian from the NW model or from the QD chain.
The self-energies are
\begin{equation}
\Sigma_0 = -\Delta \frac{\left( -E_{Z,sc} \tau_z \sigma_x + \Delta_{\text{Al}} \tau_y\sigma_y \right)}{\sqrt{\Delta_{\text{Al}}^2 - E_{Z,sc}^2}}, \quad
\Sigma_1 = -\Delta \frac{\left( \Delta_{\text{Al}}^2 \tau_0\sigma_0 - E_{Z,sc}  \Delta_{\text{Al}} \tau_x\sigma_z \right)}{\left( \Delta_{\text{Al}}^2 - E_{Z,sc}^2 \right)^{3/2}},
\end{equation}
where $E_{Z,sc} = g_\text{sc}\mu_B B/2$ is the Zeeman energy in the superconductor and $g_\text{sc}=2$ is the $g$-factor of the superconductor, $\mu_B$ is the Bohr magneton, and $B$ is the magnetic field.
Here $\Delta_{\text{Al}}$ is the superconducting gap parent superconductor.

\section{Pareto distribution of microscopic parameters}\label{app:distribution}\label{sec:pareto}
\begin{figure}[h!]
  \centering
  \includegraphics[width=0.95\linewidth]{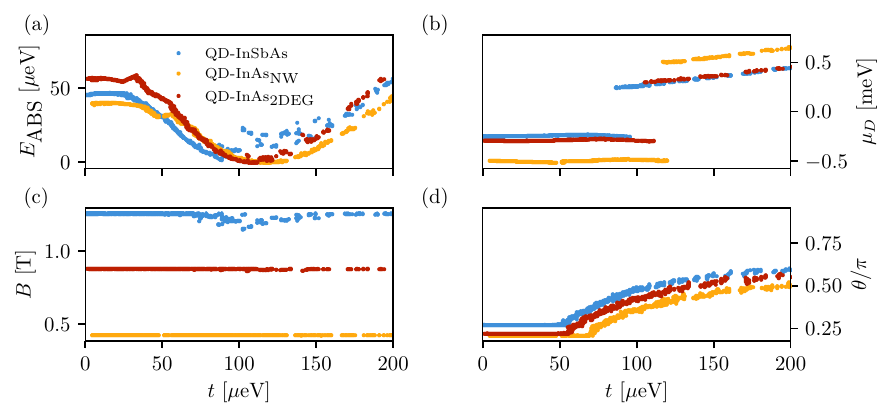}
  \caption{
    Distribution of microscopic parameters along the Pareto front for the QD chain shown in Fig.~\ref{fig:pareto} as a function of the tunel coupling $t$.
    In panels (a) and (b) we show the sublattice and QD spin expectation values as shown in Fig.~\ref{fig:topo_phase}.
    We use the QD spin expectation value as a proxy for the validity of the model.
    We choose a cutoff at $|\langle \sigma_z \tau_z \rangle_\text{QD}| > 0.8$ to ensure that the QD contains mostly one spin.
    In panels (c-f) we show the remaining microscopic parameters.
    }\label{fig:fronts_data}
\end{figure}
\comment{We set bounds on the microscopic parameters to ensure that the problem is physically sound.}
Optimising MBS in the QD chain requires to tune five different parameters.
Therefore, the Pareto front shown in Fig.~\ref{fig:pareto} describes a line in a five-dimensional space that is difficult to visualize.
To perform this optimization, we take the following considerations into account:
We impose bounds on the microscopic parameters as detailed in Table~\ref{tab:bounds}.
We limit the ABS chemical potential to ensure that the ABS remains inside the induced gap.
Because there is a finite $g$-factor in the ABS, we limit the magnetic field so that the ABS remains inside the induced gap.
Thus, the upper bound for the magnetic field is $B_\text{max} = 2(\mu_B g_\text{abs})^{-1}$.
We choose the minimal spin-orbit angle as $\theta_\text{min} = 2l_\text{dot}/l_\text{so}\pi$ where $l_\text{dot}=100$nm is the distance between quantum dots and $l_\text{so}$ is the spin-orbit length.
The bounds on the microscopic parameters are summarized in Table~\ref{tab:bounds}.
\begin{table}[h!]
    \centering
    \begin{tabular}{l|c|c|c|c|c}
        \hline
        & $|\mu_D|$ [meV] & $\mu_A$ [$\Gamma$] & $t$ [meV] & $B$ [$\Gamma$] & $\theta$ \\
        \hline
        Lower bound & -3 & 0 & 0 & 0 & $\theta_\text{min}$ \\
        Upper bound & 3 & 0.8 & 1 & $B_\text{max}$ & $\pi$ \\
        \hline
    \end{tabular}
    \caption{Bounds on the microscopic parameters used in the optimization.}\label{tab:bounds}
\end{table}

\comment{We show the distribution of points along the Pareto front for the QD chain in Fig.~\ref{fig:fronts_data}.}
To calculate the Pareto front of the QD chain, we perform two rounds of optimization~\cite{pymoo} where the first round has a positive QD chemical potential $\mu_D > 0$ and the second round has a negative QD chemical potential $\mu_D < 0$.
We combine the data from both rounds to obtain the Pareto front shown in Fig.~\ref{fig:pareto}.
In Fig.~\ref{fig:fronts_data} we analyse the microscopic parameters of the Pareto front of the QD chain as a function of the tunnel coupling amplitude $t$ for the data shown in Fig.~\ref{fig:pareto}.
In panels (a) and (b) we plot the ABS and quantum dot energies, respectively, and we observe a clear transition from $\mu_D < 0$ to $\mu_D > 0$ as the coupling $t$ increases.
This transition coincides with the point where $E_{\text{ABS}}$ reaches zero energy.
The magnetic field $B$ in panel (c) is mostly constant along the Pareto front, and it saturates at the maximum allowed value.
In panel (d) we plot the spin-orbit angle $\theta$ and observe that $\theta$ saturates the lower bound for small $t$ and then it deviates for larger values of $t$.

\begin{figure}[h!]
  \centering
  \includegraphics[width=0.5\linewidth]{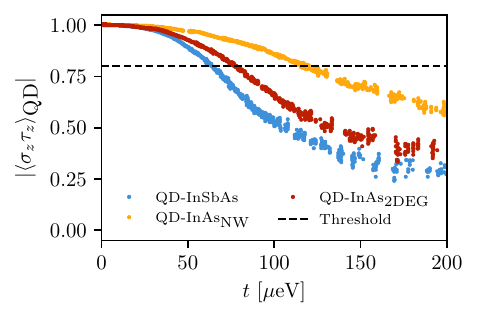}
  \caption{
    Distribution of spin expectation values along the Pareto front for the QD chain shown in Fig.~\ref{fig:pareto} as a function of the tunnel coupling $t$.
    We use the QD spin expectation value as a proxy for the validity of the model.
    We choose a cutoff at $|\langle \sigma_z \tau_z \rangle_\text{QD}| > 0.8$ to ensure that the QD contains mostly one spin so that we can disregard Coulomb interactions with the other spin.
    }\label{fig:spin_criteria}
\end{figure}

\comment{We use the QD spin expectation value as a proxy for the validity of the model.}
Our model is a valid representation of experiments as long as there is mostly one spin present in a quantum dot.
In the phase diagram of Fig.~\ref{fig:topo_phase}(a) we see that the low-energy states contain one spin when the coupling $t$ is small, but as $t$ increases, both spins become relevant.
This regime does not corresponds to the physics of the QD chain experiments~\cite{Zatelli2023Robust,Bordin2023Tunable,Bordin2024Signatures,vanLoo2025}.
Therefore, we use the QD spin expectation value $\langle \sigma_z \tau_z \rangle_\text{QD}$ as a proxy for the validity of our model.
In Fig.~\ref{fig:spin_criteria} we plot the distribution of spin expectation values along the Pareto front shown in Fig.~\ref{fig:pareto} as a function of the tunnel coupling $t$.
We choose a cutoff at $|\langle \sigma_z \tau_z \rangle_\text{QD}| > 0.8$ to ensure that the QD contains mostly one spin so that we can disregard Coulomb interactions with the other spin.
This cutoff excludes a portion of the Pareto front that has both a very large gap and large localization length.
Nevertheless, it confirms that our model captures the physics of the Kitaev limit with long-range couplings for small $t$, which is inaccessible to the NW model.

\end{document}